\documentclass[aps,twocolumn,prb,tightenlines,floatfix,showpacs]{revtex4}
\usepackage[dvips]{graphicx}
\usepackage[english]{babel}
\usepackage{amsmath}
\usepackage{amssymb}
\usepackage{times}
\newcommand{\ek}{\epsilon_{\mathbf{k}}}
\newcommand{\Ek}{E_{\mathbf{k}}}

\newcommand{\mb}[1]{{\mathbf{#1}}}

\newcommand{\phik}{\varphi_{\mathbf{k}}}

\begin{document}

\title{Understanding the protected nodes and collapse of the Fermi arcs
  in underdoped cuprate superconductors}

\author{Qijin Chen and K. Levin}

\affiliation{James Franck Institute and Department of Physics,
University of Chicago, Chicago, Illinois 60637}

\date{\today}

\begin{abstract}
  We show how recent angle resolved photoemission measurements
  addressing the Fermi arcs in the cuprates reveal a very natural
  phenomenological description of the complex superfluid
  phase. Importantly, this phenomenology is consistent with a previously
  presented microscopic theory.  By distinguishing the order parameter
  and the excitation gap, we are able to demonstrate how the collapse of
  the arcs below $T_c$ into well defined nodes is associated with the
  \emph{smooth} emergence of superconducting coherence.
  Comparison of this theory with experiment shows good semi-quantitative
  agreement.

\end{abstract}


\maketitle

In the past most of the interest in lower $T_c$ cuprate superconductors
has focused on the exotic, non-Fermi liquid normal phase.  In a recent
paper by Kanigel et al, \cite{Kanigel} angle resolved photoemission
spectroscopy (ARPES) experiments on several underdoped samples of
Bi$_2$Sr$_2$CaCu$_2$O$_8$ (Bi2212) were used to establish key features
of the superconducting phase.
In particular, it was reported \cite{Kanigel} that (i) the
ARPES-measured excitation gap, $\Delta({ \bf k},T)$ is roughly constant
in temperature from $T=0$ to above $T_c$.  (ii) Below $T_c$, $\Delta({
  \bf k})$ displays the $d$-wave point nodes which broaden into Fermi
arcs above $T_c$, with the change occurring within the width of the
resistive transition at $T_c$. (iii) It is claimed \cite{KanigelNature}
that the energy scale of the excitation gap is $T^*$, or the pseudogap
\cite{LeeReview,ourreview} onset temperature, and that the Fermi arc
length scales with $T/T^*$ above $T_c$.  From (i) it is inferred that
(iv) ``the energy gap is \textit{not} directly related to the
superconducting order parameter''.

These latest experiments have underlined the fact that the
\textit{superfluid phase} is itself very complex in the presence of a
normal state gap or ``pseudogap''. \cite{LeeReview,ourreview} Some
theories \cite{Millis,Normanarcs} seem to suggest that the already large
pseudogap becomes the order parameter immediately below $T_c$, which
would seem to imply an (unphysical) jump in the order parameter and in
the superfluid density, $n_s$.  In our approach we show how these
important photoemission observations reveal a more natural description
of the superfluid phase. We then review a microscopic model consistent
with this phenomenology which has been demonstrated \cite{ourreview} to
be compatible with a variety of other experiments and show that it
yields good semi-quantitative agreement with a large number of different
representations of these recent photoemission data.

Our microscopic scheme is based on a BCS--Bose-Einstein condensation
(BEC) crossover scenario \cite{Leggett,ourreview} and is distinct from
the phase fluctuation scenario.  \cite{Emery,TesanovicQED3,Millis} It
has been argued \cite{LeggettNature} to be appropriate to the cuprates
because of their very short coherence length.  We emphasize the generic
features of our framework. One assumes that there are attractive
interactions which lead to pairing which, in turn, gives rise to a gap
in the fermionic spectrum.  Noncondensed pairs above $T_c$ contribute to
this excitation gap, $\Delta$, just as do condensed pairs.  Thus, we
infer that below $T_c$, $\Delta$ contains two contributions,
$\Delta_{sc}$ from condensed and $\Delta_{pg}$ from noncondensed pairs.
Since the pair density is associated with the square of the gap, these
contributions add in quadrature to yield
\begin{equation}
\Delta^2 = \Delta_{sc}^2+\Delta_{pg}^2.
\label{eq:sum}
\end{equation}
The same equation is consistent with points (i) and (iv) above.  Note
that the superfluid density, $n_s(T)$ is observed to vanish smoothly as
$T_c$ is approached from below, which requires that the superconducting
order parameter $\Delta_{sc}$ turn on continuously as in a second order
phase transition.  We then conclude that because the excitation gap
$\Delta({\bf k}, T) $ is roughly a constant in $T$ across $T_c$ there
must be another component to the excitation gap below $T_c$, which
compensates for the $T$ dependence in $\Delta_{sc}(T)$. The simplest
approach is to think of this term (i.e., $\Delta_{pg}^2(T)$) as a
``fluctuation'' contribution of the form $ \langle\Delta^2\rangle -
\langle\Delta\rangle^2$.

The normal state analysis of ARPES experiments has already made
substantial use \cite{Chen4,Normanarcs,Chubukov2} of a broadened BCS
form \cite{Maly1} for the fermion self energy
\begin{equation}
  \Sigma_{pg}(\mb{k},\omega) =
  \frac{\Delta_{\mb{k},pg}^2}{\omega+\ek+i\gamma} -i\Sigma_0 \,.
\label{SigmaPG_Model_Eq}
\end{equation}
Here the broadening $\gamma \ne 0$ and ``incoherent'' background
contribution $\Sigma_0$ reflect the fact that noncondensed pairs do not lead to
\textit{true} off-diagonal long-range order. We adopt a tight binding
model for the band dispersion $\ek$, although
% $\ek = 2t (2-\cos k_x - \cos k_y)+2t_z(1-\cos k_z) -\mu$ in our
% microscopic theory.
the detailed band structure, is of no importance in our calculations
which address ARPES data points along the Fermi surface $\ek=0$.  We define
$\Delta_{\mb{k},pg}=\Delta_{pg}\phik$, and we introduce
$\varphi_{\bf k} = \cos (2\phi)$, to reflect the $d$-wave $\bf k$
dependence along the Fermi surface.  

In contrast to earlier scenarios 
\cite{Normanarcs,Millis} here we distinguish the superfluid from the
normal phases via an additional component of the self energy arising
from the condensate, $\Sigma_{sc}$ so that
\begin{eqnarray}
\Sigma({\bf k}, \omega) &=& \Sigma_{pg}({\bf k}, \omega )
+ \Sigma_{sc}({\bf k}, \omega )\\ 
\label{eq:4}
\Sigma_{sc}(\mb{k},\omega) &=& 
\frac{\Delta_{\mb{k},sc}^2}{\omega +\ek} \:,
\label{SigmaSC}
\end{eqnarray}
with 
$\Delta_{\mb{k},sc}=\Delta_{sc}\phik$.
$\Sigma_{sc}$ is associated with long-lived condensed Cooper pairs and
so it is of the same form as $\Sigma_{pg}$ but without the broadening.
With this self energy in the Green's functions, the resulting spectral
function, $A(\mb{k},\omega)=-2\,\mbox{Im}\, G(\mb{k},\omega+i0)$ can be
readily determined.  One can see
that the spectral function at all $\bf k$ contains a zero at
$\omega=-\ek$ below $T_c$, whereas it has no zero above $T_c$.  This
dramatic effect of superconducting coherence will be reflected in the
disappearance of Fermi arcs below $T_c$.

Since $\Delta_{pg}$ represents a contribution from thermal bosonic
fluctuations, it should vanish in the ground state and roughly be of the
general form
\begin{equation}
   \Delta_{pg}^2 (T) = \left(T/T_c\right)^{\nu} \Delta^2
  (T_c),  
%  \propto  n_{nc} ,
  \qquad  T \le T_c\;.
\label{eq:7}
\end{equation}
Using this equation one can determine $T_c$ as the the lowest
temperature at which $\Delta_{sc} =0$.  The exponent $\nu$ in general
varies between 3 and 3/2, which corresponds to linear and quadratic pair
dispersion, respectively.  What is important is not the value of $\nu$,
but that at $T_c$ the order parameter vanishes, and the term
$\Delta_{pg}$ accounts for the entire excitation gap $\Delta(T_c)$.

In the underdoped regime, because $\Delta(T) \approx \Delta(T_c)$, it
would appear that the smooth vanishing of the superfluid density $n_s$
at $T_c$ introduces a challenge for theory.  In one-gap scenarios
\cite{Millis,Normanarcs} [which seem to assume $\Delta_{sc}(T_c^-) =
\Delta(T_c^-)$] one needs to demonstrate explicitly why there is not a
discontinuity in $n_s$ at $T_c$.  By including two gap parameters below
$T_c$, we arrive at a natural understanding of $n_s$.  Importantly, it
is only the condensed pairs which contribute to the Meissner magnetic
screening, as expected, so that $n_s \propto \Delta_{sc}^2$, which thus
vanishes smoothly at $T_c$.  Indeed, this is a natural consequence if
one considers superconductivity as Bose condensation of Cooper
pairs. The noncondensed pairs discussed above contribute to the
destruction of $n_s$ in addition to the usual fermionic
terms.\cite{Chen2} These bosonic contributions are clearly required
since the number of fermionic excitations ($\propto T/\Delta$) is
smaller than the total fermion number for all $T \le T_c$.  The phase
fluctuation scenario \cite{Emery} similarly invokes bosonic excitations
to destroy $n_s$ at $T_c$.

We will show below that, at a semi-quantitative level, we can address
ARPES data by assuming for all $T<T^*$
\begin{equation}
1  + U  \mathop{\sum_{\bf k}}  \frac{1 - 2 f(E_{\bf k})}{2
E_{\bf k}} \varphi _{\bf k}^2 = 0,
%\qquad  T \le T_c\;,
\label{eq:gap_equation}
\end{equation}
Here $\Ek = \sqrt{ \ek ^2 + \Delta _{\bf k}^2}$, with $\Delta_{\bf k} =
\Delta \varphi _{\bf k}$; $U<0$ is the pairing interaction
strength,\cite{Chen2} and $f(x)$ is the Fermi function.  The ratio
$\Delta(T)/\Delta(0)$ as a function of $T/T^*$ is then given by a nearly
universal curve independent of the doping concentration, $x$. 
In the underdoped regime, $\Delta(T_c)$ is large so that
Eq.~(\ref{eq:gap_equation}) is consistent with a very high $T^* >>T_c$
and a roughly constant gap within the superfluid phase.

We briefly present a T-matrix approach from which Eqs.~ (\ref{eq:sum}) -
(\ref{eq:gap_equation}) were previously derived, noting that more
details can be found elsewhere. \cite{ourreview} The BCS-like constraint
in Eq.~(\ref{eq:gap_equation}) can be interpreted as equivalent to the
BEC condition that the noncondensed pairs have zero chemical potential
%\begin{equation}
$\mu_{pair} = 0$.
%\label{eq:9}
%\end{equation}
at and below $T_c$. This determines the form of the noncondensed pair
propagator $t_{pg}(Q)= U/ [1+U \chi(Q)]$ such that $t_{pg}(0) = \infty$
at $T \leq T_c$, where
$\chi(Q)=\sum_{K}G_{0}(Q-K)G(K)\varphi_{\bf k-q/2}^2$,
is the pair susceptibility.  Here $G$ and $G_0$ are the full and bare
Green's functions, respectively.  To make direct association with
Eq.~(\ref{eq:gap_equation}), we drop the $\gamma$ and $\Sigma_0$ term
(which are more important above than below $T_c$) in
Eq.~(\ref{SigmaPG_Model_Eq}) and arrive at
$\Sigma({\bf k}, \omega ) \approx \Delta_{\bf k}^2/(\omega
+\epsilon_{\bf k} ) $ for $ T \le T_c$.  This reasonable approximation
can be used to determine the form for $G$, and establish an equivalence
between Eq.~(\ref{eq:gap_equation}) and the BEC condition
$\mu_{pair}=0$.

\begin{figure}
\centerline{\includegraphics[width=3.3in,clip]{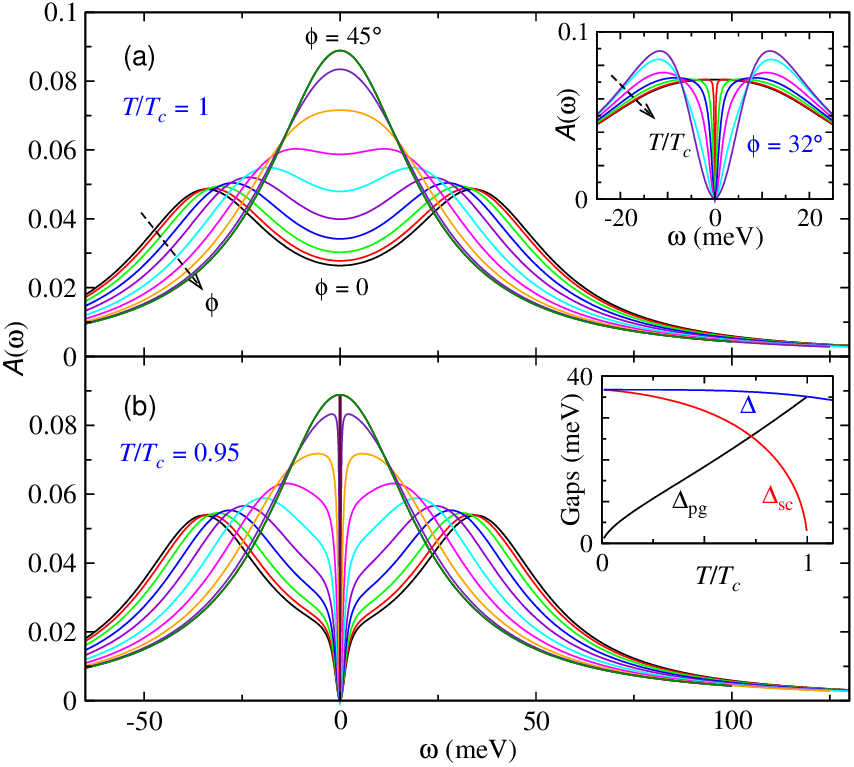}}
%{SpDf0.875r300g0.378gm0.0906Sg0.09R0.009_Fig1c.eps}
\caption{(color online) Spectral function $A(\omega)$ for $x=0.125$ at
(a) $T=T_c$ and (b) $0.95T_c$ at different angles $\phi$ along the
Fermi surface of a cuprate superconductor, where $\phi$ increases by
$4.5^\circ$ from 0 to $45^\circ$ along the direction of the arrow. The
upper inset shows the temperature dependence of $A(\omega)$ at
$\phi=32^\circ$ near the node, with $T/T_c=1$, 0.99, 0.95, 0.9, 0.8,
0.6 and 0.3 in the direction of the arrow. Shown in the lower inset
are the gaps vs $T$ below $T_c$. The effects of phase coherence are
more pronounced in the nodal region of the Fermi surface, where the
gap is small.  }
\label{fig:1}
\end{figure}

To establish the validity of Eqs.~(\ref{SigmaPG_Model_Eq}) and
(\ref{SigmaSC}), we note that there are two contributions to the full
$T$-matrix $t = t_{pg} + t_{sc}$ where the condensate contribution
$t_{sc}(Q)= -\frac{\Delta_{sc}^2}{T} \delta(Q)$.
Similarly, we have for the fermion self energy
$\Sigma (K) = \Sigma_{sc}(K) + \Sigma_{pg} (K) =
\sum_Q t(Q) G_{0} (Q-K)\varphi_{\bf k-q/2}^2$, from which
Eq.~(\ref{SigmaSC}) follows at once.
A vanishing chemical potential means that $t_{pg}(Q)$ diverges at $Q=0$
when $T\le T_c$. Thus, in the superfluid phase only, we approximate
\cite{Maly1,Kosztin1} $\Sigma_{pg}(K)$ to yield
$\Sigma_{pg} (K)\approx -G_{0} (-K) \Delta_{\mathbf{k},pg}^2 $,
where $\Delta_{pg}^2 \equiv -\sum_{Q\neq 0} t_{pg}(Q)$.
We thus may write $\Sigma_{pg} ({\bf k}, \omega) =
\Delta_{\mathbf{k},pg}^2(T)/(\omega +\epsilon_{\bf k})
+\text{small~corrections} $, where we accommodate the corrections
\cite{Maly1} with the broadening factor $\gamma$ and additional term
$\Sigma_0$.  This is necessary in order to address the \textit{concrete}
fermion spectral function.
In this way, Eq.~(\ref{SigmaPG_Model_Eq}) then follows.
Finally, at small four-vector $Q \equiv(\Omega, \mathbf{q})$, we expand
the inverse of $t_{pg}$, after analytical continuation, to obtain a
simple quadratic dispersion \cite{ourreview} for the pairs
$\Omega_{\mathbf{q}} \approx q^2/(2 M^*), $ implying $\nu = 3/2$.

To illustrate the simple physics, we do \emph{not} attempt to do
detailed curve fitting. We use one set of parameters for all doping $x$.
Thus, in our numerical results we take only $\Sigma_0$ and $\gamma
(95K)$ as adjustable to optimize overall fits to the multiple data sets.
More concretely, we obtain the universal curve $\Delta(T)/\Delta(0)$ as
a function of $T/T^*$ from Eq.~(\ref{eq:gap_equation}), say, at optimal
doping, and use the experimentally known value of $\Delta(0)$ at given
doping concentration $x$ to determine an input $T^*$ and input gap
$\Delta(T)$.  This value of $T^*$ is to be distinguished from the $T^*$
obtained in ARPES experiments, here called $T^*_{\text{ex}}$.  For our tight
binding, quasi two-dimensional lattice, Eq.~(\ref{eq:gap_equation})
yields $2 \Delta(0)/T^* \approx 4.3 $.  In ARPES experiments, $T^*_{\text{ex}}$
is determined as the temperature where $\gamma$ and $\Delta$ are roughly
equal so that there is no observable density depletion at the Fermi
level. The experimentally deduced ratio from ARPES data
\cite{KanigelNature} is slightly larger than 5 for moderately underdoped
samples, and increases with underdoping. A higher ratio of 8 has been
observed in \emph{local} STM measurements. \cite{Yazdani2} The next step
theoretically is to use the known $T_c$ to determine $\Delta_{pg}(T)$
and $\Delta_{sc}(T)$ below $T_c$, using Eqs.~(\ref{eq:7}) and
(\ref{eq:sum}). One can then determine an ``experimental'' $T^*_{\text{ex}}$
and spectral gap following the ARPES prescription for given parameters
$\gamma$ and $\Sigma_0$.

To make this determination, we choose our adjustable parameters
$\Sigma_0$ and $\gamma$ via a rough fit to ARPES data near $T_c (\approx
95\mbox{K})$ at optimal doping. Since $\Sigma_0$ is primarily governed by the
particle-hole channel, \cite{Chen4} we take it to be independent of
doping and $T$, and given by $\Sigma_0 = 26$ meV.  The broadening
parameter $\gamma$ depends primarily on temperature. Consistent with
scattering rate measurements in the literature,\cite{Hardy1,Valla2} we
take $\gamma=26$~meV at 95~K, with $\gamma = \gamma(95\mbox{K})
(T/95\mbox{K}) $ above $T_c$ and $\gamma=\gamma(T_c) (T/T_c)^3$ below
$T_c $ for given doping $x$, as used
earlier.\cite{powerlawfootnote}. Finally, this yields a ratio
$2\Delta(0)/T^*_{\text{ex}} = 5.6 \sim 5.7$ for the cases presented below,
consistent with experiment.\cite{KanigelNature} Throughout $\phi=0$ and
$\pi/4=45^\circ$ denote the anti-nodal and nodal directions,
respectively. In order to compare with ARPES data, we convolve the
spectral function with a Gaussian instrumental broadening curve with a
standard deviation $\sigma=3$~meV, given by the ideal ARPES resolution.

In the inset to Fig.~\ref{fig:1}(b) we plot the various gap parameters
$\Delta_{sc}$, $\Delta$ and $\Delta_{pg}$ as a function of temperature
below $T_c$ and slightly above, for a typical underdoped system.  Here
for definiteness we have chosen $\nu = 3/2$ in Eq.~(\ref{eq:7}). It can
be seen that, as $T$ is lowered below $T_c$, the total excitation gap
$\Delta$ is essentially a constant while $\Delta_{sc}$ increases from 0
at $T_c$ to reach the full gap value at $T=0$. By contrast,
$\Delta_{pg}$ monotonically decreases to 0.

\begin{figure}
\centerline{  \includegraphics[width=3.3in,clip]{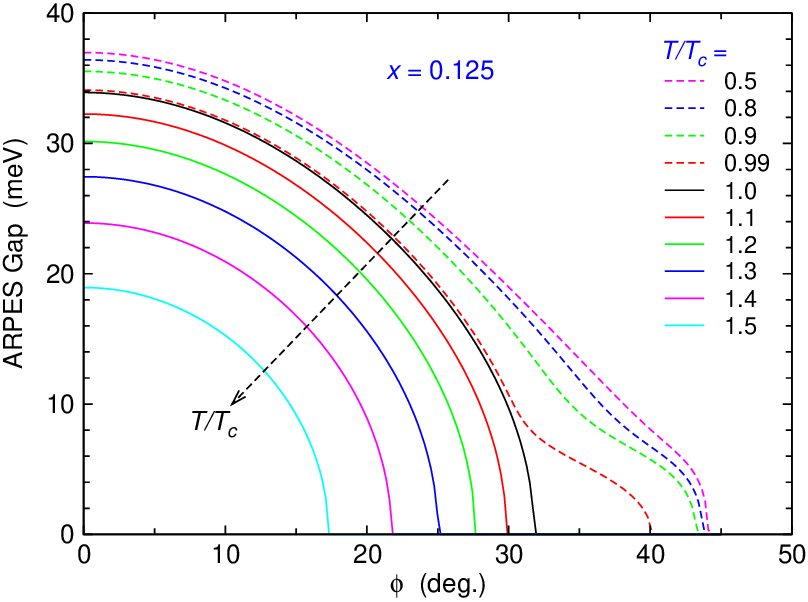} 
%  {ARPESGapDf0.875r300g0.378Sig0.1044gm0.105R0.0104-phi-T_Fig2.eps}
}
\caption{(color online) Spectral gap as measured in ARPES as a function
  of angle $\phi$ along the Fermi surface of an underdoped cuprate
  superconductor of doping $x=0.125$, at different $T/T_c$ between 0.5
  and 1.5.  The spectral function has been broadened with a small ARPES
  instrumental resolution of 3 meV. A Fermi arc appears as the extended
  range of zero gap value. The fast departure of the $T=0.99T_c$ curve
  from the $T=T_c$ curve indicates a collapse of the Fermi arc.}
\label{fig:2}
\end{figure}

In Figs.~\ref{fig:1}(a) and \ref{fig:1}(b), we plot the spectral
function $A(\omega)$ for $\ek=0$ (on the Fermi surface) at and slightly
below $T_c$, respectively, and for different angles. To illustrate the
physics, here we do not include the ARPES instrumental broadening.
Just below $T_c$, the sharp dip at $\omega =0$ is associated with the
onset of a very small condensate, which nevertheless leads to a
depletion of the spectral weight at the Fermi level.  The coherence
associated with the order parameter is better illustrated near the nodal
region where a gap, absent at $T_c$, appears as $T$ decreases, as shown
for $\phi= 32^\circ$ in the upper inset.  A closer inspection of the
shape of $A(\omega)$ in Fig.~\ref{fig:1}(b) suggests that $A(\omega)$
cannot be described by a simple broadened BCS spectral function.

\begin{figure}
\centerline{\includegraphics[width=3.3in,clip]{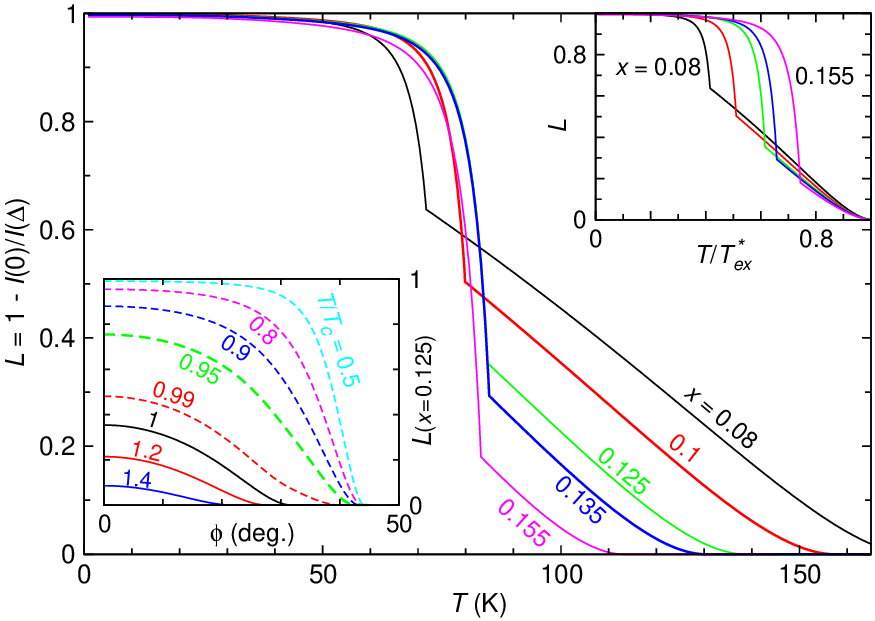}}
%{Lphi0Dg0.315lam0Sig0.087gm0.087R0.0087-T-x_Fig3.eps}
\caption{(color online) Loss of zero energy spectral weight as a
  function of temperature at the antinode, for different hole
  concentrations.  Upper right inset shows rescaled $T$ dependence,
  while lower left inset indicates $\phi$ dependence for various $T$.}
\label{fig:3}
\end{figure}

Figure \ref{fig:2} presents a plot of the spectroscopic gap for an
underdoped cuprate with $x=0.125$ as a function of angle and for various
temperatures below (dashed curves) and above $T_c$ (solid curves).  Here
the spectral gap is given by half the peak-to-peak distance in the
spectral function, and is smaller than the input gap $\Delta(T)$.
For $T<T_c$, since the gap is roughly $T$ independent, the various
curves tend to coalesce. Above $T_c$, the extended range of zero gap
value around $\phi=45^\circ$ gives rise to the Fermi arcs.
\cite{Normanarcs,Chubukov2} This is due to the presence of $\gamma$. The
rapid deviation of the (red dashed) $T=0.99T_c$ curve from the (black
solid) $T=T_c$ curve indicates a collapse of the Fermi arc, leading to
the protected nodes below $T_c$, reflecting the emergence of
superconducting order below $T_c$. This figure should be compared with
Fig.~2 of Ref.~\onlinecite{Kanigel}.

In Fig.~\ref{fig:3} we plot the relative loss of spectral weight
$L(\phi)= 1 - I(0)/I(\Delta_{\phi})$, defined following
Ref.~\onlinecite{Kanigel}, at the anti-node $\phi=0$ as a function of
$T$ for different doping concentrations, where $I(\omega)\equiv
A(\omega)$. This determines $T^*_{\text{ex}}$ as where $L(0)$ vanishes.
The lower left inset presents plots of $L(\phi)$ as a function of $\phi$
for various temperatures while the upper right inset illustrate the
anti-node behavior $L(\phi=0)$ as a function of $T/T^*_{\text{ex}}$.
Above $T_c$ the dependence is linear in $T$ reflecting a linear $T$
dependence in $\gamma$. Illustrated in the upper inset is a considerable
universality above $T_c$ as a function of $T/T^*_{\text{ex}}$, which
provides a prediction for future ARPES data analysis. As the temperature
decreases, the abrupt jump at $T_c$ reflects the onset of phase
coherence, as in Fig.~\ref{fig:2}.  Here the $x=0.1$ and $x=0.135$ lines
are close to the data for the 67K and 80K samples in the upper panel of
Fig.~4 in Ref.~\onlinecite{Kanigel}.

\begin{figure}
 \centerline{ \includegraphics[width=3.3in,clip]{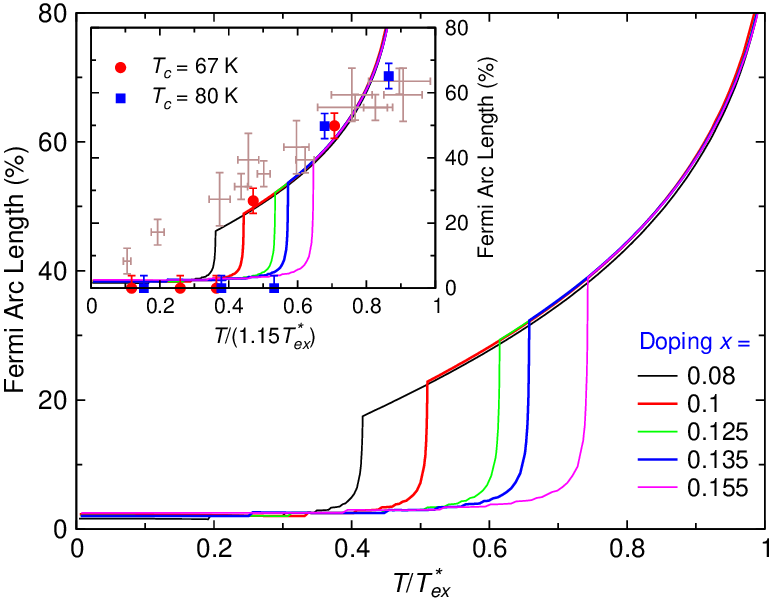}
%{FALeng.315r300lam0Sig0.087gm0.087R0.0087-T-x_Fig4_3.eps}
}
\caption{(color online) Fermi arc length as a function of
  $T/T^*_{\text{ex}}$ for doping concentrations from optimal to
  underdoping for a cuprate superconductor. Fermi arc length is
  typically finite above $T_c$ and drops to zero upon the onset of phase
  coherence. The normal state portions of the curves is close to
  universal, in agreement with Ref.~\onlinecite{KanigelNature}.  The
  comparison in the inset between the theory with a slightly (15\%)
  enlarged $T^*_{\text{ex}}$ and experimental data (symbols)
  \cite{Kanigel} shows a good semi-quantitative agreement.}
\label{fig:4}
\end{figure}

Finally, Fig.~\ref{fig:4} shows the sharp collapse of the Fermi arcs
from above to below $T_c$; we plot the percentage of arc length as a
function of $T/T^*_{\text{ex}}$ and for different doping concentrations
from the optimal to the underdoped regime. The small nonvanishing arc
length at low $T$ reflects the finite ARPES resolution.
As a consequence of the Fermi arc collapse below $T_c$, the nodes are
``protected''.  In addition, there is a clear universality seen in the
normal state, in good agreement with the central features observed
experimentally, shown in Fig.~4 of Refs.~\onlinecite{Kanigel} and
\onlinecite{KanigelNature}. Since the mean-field equation is only a
crude approximation above $T_c$, it is reasonable to allow
$T^*_{\text{ex}}$ to vary slightly, as we have done here, to compare
semi-quantitatively with the data. By contrast with
Ref.~\onlinecite{Kanigel}, however, our curves are not strictly straight
lines, reflecting the nonlinearity of the gap as a function of
$\phi-\pi/4$, as also found in Ref.~\onlinecite{Normanarcs}; this seems
to be consistent with Fig.~4 of Ref.~\onlinecite{KanigelNature}.

The microscopic approach \cite{ourreview} considered here is associated
with stronger-than-BCS attractive interactions which lead to small pair
size. A major consequence of this theory (as well as the equivalent
phenomenology which ARPES experiments lead us to infer) is that
\textit{pseudogap effects persist below $T_c$ in the form of
  noncondensed pair excitations of the condensate}.  Importantly, this
leads to two contributions to the self energy below $T_c$ [see
Eq.~(\ref{eq:4})].  We find that the collapse of the Fermi arcs is not
to be associated with an abrupt disappearance (as assumed elsewhere
\cite{Normanarcs}) of the inverse pair lifetime $\gamma$, appearing in
$\Sigma_{pg}$, but rather it reflects the \emph{gradual} emergence of
the condensate, appearing in $\Sigma_{sc}$, to which the finite momentum
pairs are continuously converted as $T$ decreases.

This work was supported by NSF Grants No. PHY-0555325 and No. MRSEC
DMR-0213745.

\bibliographystyle{apsrev}

%\bibliography{Review2}

\begin{thebibliography}{19}
\bibitem[{pow()}]{powerlawfootnote}
\bibinfo{note}{See footnote 42 of Ref.~\onlinecite{Chen4}. We have also checked
  that below $T_c$ the Fermi arc length is not sensitive to the power law of
  $\gamma(T)$.}
\end{thebibliography}

\end{document}